\documentclass[prl,twocolumn]{revtex4-1}

\usepackage{amsmath}
\usepackage{amssymb}
\usepackage{amsthm}

\usepackage{graphicx}
\usepackage{graphics}
\usepackage{bbold,bm}
\usepackage{hyperref}
\usepackage{epsfig}
\usepackage{dcolumn}

\def\bra#1{\mathinner{\langle{#1}|}}
\def\ket#1{\mathinner{|{#1}\rangle}}
\def\braket#1{\mathinner{\langle{#1}\rangle}}

\def\bbra#1{\mathinner{\langle\hspace{-0.75mm}\langle{#1}|}}
\def\kket#1{\mathinner{|{#1}\rangle\hspace{-0.75mm}\rangle}}
\def\bbraket#1{\mathinner{\langle\hspace{-0.75mm}\langle{#1}\rangle\hspace{-0.75mm}\rangle}}

\def\re{\mathrm{Re}\,}
\def\im{\mathrm{Im}\,}
\def\tr{\mathrm{Tr}}
\def\dd{\mathrm{d}}

\def\ii{i}
 
\def\dbar{\hbox{$d$\kern-0.6em\raise0.3em\hbox{$-$}}\hspace{-0.5mm}}

\begin{document}

\title{Transport signatures of Floquet Majorana fermions in driven topological superconductors}

\author{Arijit Kundu}

\address{Department of Physics, Indiana University, 727 East Third Street, Bloomington, IN 47405-7105 USA}

\author{Babak Seradjeh}

\address{Department of Physics, Indiana University, 727 East Third Street, Bloomington, IN 47405-7105 USA}

\begin{abstract}
Floquet Majorana fermions are steady states of equal superposition of electrons and holes in a periodically driven superconductor. We study the experimental signatures of Floquet Majorana fermions in transport measurements and show, both analytically and numerically, that their presence is signaled by a
quantized conductance sum rule over discrete values of lead bias
differing by multiple absorption or emission energies at drive frequency. We also study
the effects of static disorder and find that the quantized sum rule is robust against weak disorder. Thus, we offer a unique way to identify the topological 
signatures of Floquet Majorana fermions.
\end{abstract}

\maketitle

\emph{Introduction.}---The nonlocal quantum order characterizing the topological state of a gapped medium often necessitates the existence of topologically protected gapless states bound to bulk defects or the edge with a topologically trivial medium where the gap closes. The detection of these topological bound states is, therefore, a primary probe of the topological state.
It has recently been understood that topological bound states may arise as steady states when a topologically trivial system is driven periodically~\cite{OkaAok09a,LinRefGal11a,KitOkaBra11a}. In the
superconducting state, these are equal superpositions of electrons and holes
known as Floquet Majorana fermions~\cite{JiaKitAli11a,LiuHaoZhu12a} that exhibit non-Abelian statistics~\cite{Iva01a,LiuLevBar12a} and can be used for topological quantum computation~\cite{Kit03a}.  This possibility expands the systems and conditions that realize Majorana fermions as emergent quasiparticles~\cite{Kit01a,FuKan08a,SauLutTew10a,Ali10a,LutSauDas10a,OreRefOpp10a,Wil09a,Ali12a,Bee11a}, but also poses fundamental questions as to how to detect and possibly manipulate such steady states. In particular, since the driven system is not in equilibrium, the experiments probing the equilibrium response of static Majorana fermions~\cite{MouZuoFro12a,DasRonMos12b,WilBesGal12b,RokLiuFur12a,FinVanMoh13a} cannot be used directly for this purpose.

In this Letter, we address these questions by studying the non-equilibrium transport properties of Floquet Majorana fermions. We show, both analytically and numerically, that there is a quantized conductance sum rule, which we dub the ``Floquet sum rule,'' whenever Floquet Majorana fermions exist. The Floquet sum rule naturally generalizes the quantized zero-bias conductance of static Majorana fermions~\cite{BolDem07a,NilAkhBee08a,LawLeeNg09a,Fle10a,HutZazBra12b,LiuPotLaw12b}.  
Moreover, we show that the Floquet sum rule is robust against moderate static disorder, owing to its topological character, while other peaks get suppressed. Remarkably, this suggests that disorder, usually detrimental to electronic properties, can be used as a ``sieve'' to find Floquet Majorana fermions. Transport studies in irradiated graphene, where the Floquet topological insulator was first proposed to exist~\cite{OkaAok09a}, suggested quantized transport in the driven system is possible in certain geometries and for large drive frequencies~\cite{KitBerRud10a,GuFerAro11a}. 
We use a systematic Green's function method that extends the previous studies to superconducting systems in any frequency range, and can, in principle, incorporate the effects of interactions.

Though our results are applicable to any realization of Floquet Majorana fermions, systems of cold atoms could prove specially useful in this regard due to a high degree of design control and newly developed experimental probes of their dynamics, such as single-atom imaging, tunneling, and transport~\cite{KolKohGia07a,BakGilPen09a,SheWeiEnd10a,BruBel12a,BraMeiSta12a,StaKriMei12a,ChiVen12a}. Disorder can be introduced in cold atom systems controllably~\cite{LyeFalMod05a,PasMcKWhi10a} and could, therefore, play a key role in the detection and manipulation of Floquet Majorana fermions. In the solid state, such as in quantum wires~\cite{Kit01a,LutSauDas10a,OreRefOpp10a,MouZuoFro12a,DasRonMos12b,FinVanMoh13a}, high-frequency irradiation of the order of the bandwidth is detrimental to the proximity-induced superconducting state. However, we find numerically that even at much lower frequencies, Floquet Majorana fermions can still be realized and have the same transport signatures, with or without disorder, as at higher frequencies.

\emph{Model.}---We study the model Hamiltonian $H(t)=H_{\mathrm{w}}(t)+H_{\mathrm{c}}+H_{\mathrm{l}}$, where the last term describes the leads,
\begin{equation}
H_{\mathrm{w}}(t) = \frac{\ii}2 \gamma^{\intercal}A(t)\gamma,
\end{equation}
is the Hamiltonian of the system (wire) in the Majorana basis $\gamma^{\intercal}=(\gamma_{1},\cdots,\gamma_{2N})$ with a real, skew-symmetric matrix $A(t)$, and the contact Hamiltonian $H_{\mathrm{c}}=\sum_{\lambda}a^{\lambda\dagger}K^{\lambda}\gamma+\mathrm{h.c.}$ with $a^{\lambda\dagger}$ is the row of electronic creation operators in lead $\lambda$, and $K^{\lambda}$ a contact matrix. 

Our analytical results are presented for a general realization of Majorana fermions. For numerical calculations, we choose the simple model of a single-band quantum wire with superconducting pairing in a spin-polarized electronic band~\cite{Kit01a,SerGro11a}. This model can be effectively realized in solid state~\cite{LutSauDas10a,OreRefOpp10a,MouZuoFro12a,DasRonMos12b,FinVanMoh13a} and potentially in cold atom systems~\cite{JiaKitAli11a,HoZha11a,WanYuFu12a,CheSomHad12a,LiuDru12a}.
There are two Majorana operators $(\gamma_{r1},\gamma_{r2})\equiv\gamma_{r}^{\intercal}$ at sites $r=1,\cdots L$.
The contact matrix elements $K^\lambda \propto (1,i)$ in the Majorana basis  at each site. 
The nonzero elements of $A$ are 
\begin{equation}
A_{r,r}=-\ii \mu_{r} \sigma_{y},\quad
A_{r,r+1}= \Delta_{r} \sigma_{x} + \ii\, w_{r} \sigma_{y},
\end{equation}
and $A_{r+1,r}=-A_{r,r+1}$, where the real parameters $\mu_{r}$, $\Delta_{r}$ and $w_{r}$ are, respectively, the chemical potential at site $r$, the superconducting pairing, and the hopping integral on $(r,r+1)$ bond, and $(\sigma_{x},\sigma_{y},\sigma_{z})$ are Pauli matrices. The static, uniform wire with bandwidth $W=4|\re w|$ has a topological phase transition at $2|\mu|=W$ where the gap closes. There is an unpaired Majorana fermion at each end of the wire with zero energy (the energy is referenced to chemical potential of the wire) in the topological phase $2|\mu|<W$ and none in the trivial phase $2|\mu|>W$~\cite{Kit01a}.

\begin{figure}[t]
\includegraphics[scale=1]{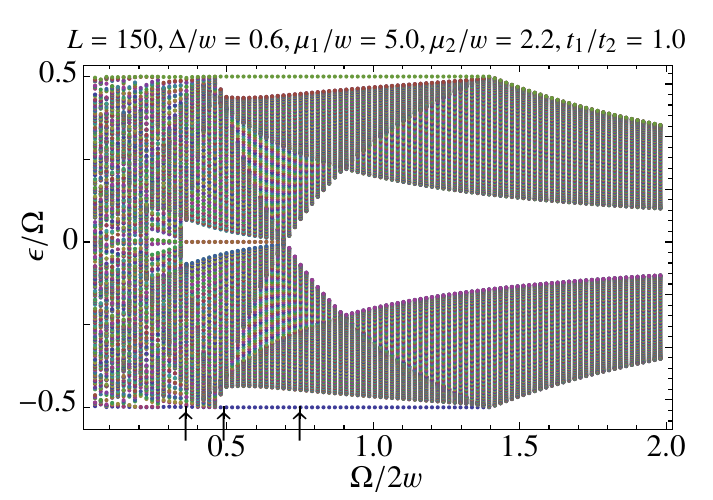}
\caption{The evolution of the Floquet spectrum of a quantum wire with $L$ sites vs. $\Omega$ for a square-wave chemical potential. The arrows show the frequencies used for the transport calculations in Fig.~\ref{fig:floquet}.}\label{fig:floqspec}
\end{figure}

When the system is driven with period $T\equiv2\pi/\Omega$, the general solution of the time-periodic Schr\"odinger equation (we set $\hbar=1$) $H(t)\ket{\psi(t)}= \ii\partial_{t}\ket{\psi(t)}$ is found in terms of the Floquet functions $\ket{\psi_{\alpha}(t)} = e^{-\ii\epsilon_{\alpha}t}\ket{\phi_{\alpha}(t)}$, where $\ket{\phi_{\alpha}(t+T)}=\ket{\phi_{\alpha}(t)}$ is an eigenket of the effective Hamiltonian $H_{\mathrm{eff}}(t)=H(t)-\ii\partial_{t}$, $H_{\mathrm{eff}}(t)\ket{\phi_{\alpha}(t)}=\epsilon_{\alpha}\ket{\phi_{\alpha}(t)}$. The quasienergies $\epsilon_{\alpha}$ are restricted to $(-\Omega/2,\Omega/2]$ by the map $\epsilon_{\alpha}\mapsto\epsilon_{\alpha}+k\Omega$, $\ket{\phi_{\alpha}(t)}\mapsto e^{\ii k\Omega t}\ket{\phi_{\alpha}(t)}$. We shall compute the Floquet spectrum $\{\epsilon\}$ using the evolution operator $U(t)\ket{\psi(0)}=\ket{\psi(t)}$ and constructing the Floquet Hamiltonian $H_{F}=(i/T)\log[U(T)]$. Then, $H_{F}\ket{\phi_{\alpha}(0)}=\epsilon_{\alpha}\ket{\phi_{\alpha}(0)}$. The periodic eigenkets can be resolved in a Fourier series $\ket{\phi(t)}=\sum_k e^{-\ii k\Omega t}\ket{\phi^{(k)}}$. We shall use a shorthand $\kket{\phi_\alpha}$ for vectors in the extended Hilbert space spanned by $\ket{\phi_\alpha^{(k)}}$, with the inner product $\bbraket{\phi'|\phi}\equiv\sum_k\braket{\phi'^{(k)}|\phi^{(k)}}=\int_0^T\braket{\phi'(t)|\phi(t)}\dd t/T$~\cite{Sam73a}.

\emph{Floquet Majorana fermions.}---Floquet Majorana fermions are bound states with quasienergy $\epsilon_0=0$ or $\epsilon_\pi=\Omega/2$~\cite{JiaKitAli11a}. The particle-hole symmetry, $H_{\mathrm{w}}^\intercal=-H_{\mathrm{w}}$, requires $H_F^\intercal=-H_F$, so the quasienergies come in pairs $(\epsilon_\alpha,-\epsilon_\alpha)$. In the Nambu basis, $\frac1{\sqrt 2}\left(\begin{array}{cc} 1 & \ii \\ 1 & -\ii\end{array}\right)\phi_\alpha=\left(\begin{array}{c}u_{\alpha} \\ v_{\alpha}\end{array}\right)$, the $\epsilon_0$ and $\epsilon_\pi$ Floquet Majorana fermions $v_{0}(t)=u_{0}^*(t)$ and $v_{\pi}(t)=e^{\ii\Omega t}u_{\pi}^*(t)$.

In Fig.~\ref{fig:floqspec}, we show the Floquet spectrum of a wire with a square-wave periodic chemical potential, $\mu(t)$, alternating with frequency $\Omega$ between $\mu_1$ and $\mu_2$, respectively, over time intervals $t_1$ and $t_2$ in each period. Note that $\mu(t)$ is not in the topological range at any time. As $\Omega/W$ decreases, the Floquet band spreads and $\epsilon$ crosses $\epsilon_\pi$, giving rise to a pair of $\epsilon_\pi$ Floquet Majorana fermion bound states. Additional gap-closing level crossings lead to the appearance or disappearance of $\epsilon_0$ and $\epsilon_\pi$ Floquet Majorana fermions. 

A high-frequency approximation for $\Omega\gg W$ can be made using the Baker-Campbell-Hausdorff formula, $T H_F=t_1 H_1 + t_2 H_2 + t_1t_2 [H_2,H_1] + \frac1{12}t_1t_2 [t_2H_2-t_1H_1,[H_2,H_1]]+ \cdots$. The first two terms yield a static quantum wire with an averaged chemical potential $\bbraket{\mu}=(t_1\mu_1+t_2\mu_2)/T$. The only terms contributing to the commutator $[H_2,H_1]$ are $\Delta\sigma_x$ in $A_{r,r+1}$ and $-\ii\mu\sigma_y$ in $A_{r,r}$, yielding a $\sigma_z$ term in $A_{r,r+1}$ that contributes to $\im w$. Physically, $\im w$ introduces a supercurrent in the chain, which renormalizes the spectral gap but, when small, leaves the topological phase boundary unchanged~\cite{SerGro11a}. The same is true for the next term shown. Therefore, when $2\bbraket{|\mu|}>W$, there are no Floquet Majorana fermions in the high-frequency limit. In the low-frequency limit $\Omega\ll W$ multiple exchange processes with energy $\Omega$ become important and result in qualitative differences between the static energy and quasienergy spectra. We note here that, as can be seen in Fig.~\ref{fig:floqspec}, Floquet Majorana fermions are found numerically for a much wider range of parameters, including $\Omega\ll W$ and $2\bbraket{|\mu|}>W$~\cite{LiuLevBar12a,TonAnGon12a}. 


\emph{Transport.}---Electrons in a driven system do not follow the usual statistics in a closed system. In the transport problem we can address this issue by assuming that the leads are static and follow the usual Fermi-Dirac statistics at the distant past. The scattering problem between the leads can then be formulated by integrating out the leads using their Green's function~\cite{KohLehHan05a,Note1}. This procedure adds to $H_{\mathrm{w}}$ an imaginary self-energy $\ii\Gamma(t)=\ii\sum_\lambda \Gamma^\lambda(t)$, where
$
\Gamma^\lambda(t)=2\im\left[ K^{\lambda\dagger}g^\lambda(t)K^\lambda\right],
$
and $g^\lambda$ is the Green's function of lead $\lambda$.
The wire's Green's function is periodic, $G(t+T,t'+T)=G(t,t')$, and satisfies
\begin{equation}
[\partial_t-A(t)]G(t,t')-(\Gamma*G)(t,t') =-\ii\delta(t-t'),
\end{equation}
where $\Gamma*G$ is the convolution $\int_{-\infty}^t\Gamma(t-s)G(s,t')\dd s$. Then the steady state (time-averaged) current in lead $\lambda$, $J^\lambda=\ii e\bbraket{[H(t),a^{\lambda\dagger}a^\lambda]}$, can be computed with the Green's function. Assuming the leads' density of states $\rho^\lambda$ is constant over the scattering energy range, we find an energy-independent $\Gamma^\lambda=-(\xi^\lambda+\xi^{\lambda\intercal})/2$, $\xi^\lambda=2\pi K^{\lambda\dagger}\rho^\lambda K^\lambda$ and the differential conductance $\sigma^\lambda=\dd J^\lambda/\dd V_\lambda$, with bias $V_\lambda$, reads
\begin{align}
\sigma^\lambda &={}
 -\frac{e^2}{2\pi} \int\dd\omega \bigg[  \sum_{n} \left\{ \tr[{\xi^\lambda}^{\intercal} G^{(n)}(\omega)\xi^{\lambda} G^{(n)\dagger}(\omega)] f_\lambda'(\omega)  \right. \nonumber \\
&+  \left.
\tr[\xi^\lambda G^{(n)}(\omega){\xi^\lambda}^{\intercal} G^{(n)\dagger}(\omega)] f_\lambda'(-\omega) \right\} + C_\lambda(\omega) f'_\lambda(\omega) \bigg]
\end{align}
where $f_\lambda(\omega)=[1+e^{(\omega-e V_\lambda)/\tau_\lambda}]^{-1}$ is  the Fermi-Dirac distribution, $G^{(n)}(\omega)=\frac1T\int_0^T\int e^{\ii n\Omega t}e^{\ii\omega s}G(t,t-s)\dd s\dd t$, and $C_\lambda(\omega) =  \sum_{\kappa\neq\lambda,n}\tr[(\xi^\kappa+{\xi^\kappa}^{\intercal})G^{(n)}(\omega)\xi^\lambda G^{(n)\dagger}(\omega)]$~\cite{Note1}.
The static case is found by setting $G^{(n)}(\omega)=\delta_{n,0}G(\omega)$. 

For a single lead with a point contact, $C_\lambda$ vanishes identically and the contact matrix is zero except at the contact site where $\xi=\nu(1-\sigma_y)/2$ with $\nu=2\pi\rho|w|^2$. Then,
\begin{equation}
\sigma = -\frac{e^2\nu^2}{2\pi}\sum_n\int\left[|\mathcal{G}_{\mathrm{he}}^{(n)}(\omega)|^2+|\mathcal{G}_{\mathrm{eh}}^{(n)}(-\omega)|^2\right]f'(\omega)\dd\omega,
\end{equation}
where the $\mathcal{G}_{\mathrm{eh}}$ and $\mathcal{G}_{\mathrm{he}}$ are the off-diagonal elements of the Nambu-Gorkov Green's function at the contact site,
\begin{equation}
\mathcal{G}^{(n)}(\omega)=\sum_{\alpha k} \frac{\ket{\varphi_\alpha^{(n+k)}}\bra{\bar{\varphi}_\alpha^{(k)}}}{\omega-\epsilon_\alpha-n\Omega+\ii \delta_\alpha},
\end{equation}
with $-i\delta_\alpha$ the self-energy correction to quasienergy $\epsilon_\alpha$, and $\ket{\varphi_\alpha}$ and $\bra{\bar{\varphi}_\alpha}$, respectively, the right and left Floquet eigenvectors of the effective (non-Hermitian) Hamiltonian $H_{\mathrm{w}}+\ii\Gamma$ at level $\alpha$.

\begin{figure}[t]
\includegraphics[width=2.6in]{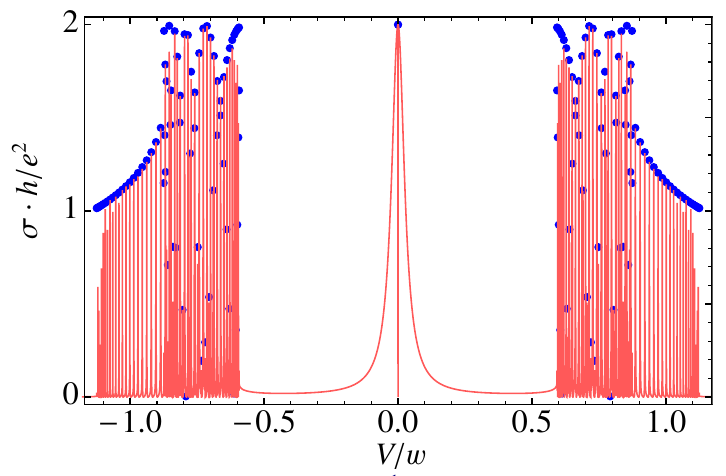}
\caption{Differential conductance $\sigma=\dd I/\dd V$ vs. bias $V$ in a single-terminal setup for a static system with $L=80, \Delta/w=0.6, \mu/w=0.25$ and $\nu/w=\pi/25$. The (blue) dots are calculated from the analytical expression of peak heights.}\label{fig:static}
\end{figure}

In the weak-contact limit $\nu/w\ll1$, we can employ perturbation theory in $\Gamma$. To the leading order, we find $\ket{\varphi_\alpha}=\ket{\phi_\alpha}$, $\bra{\bar\varphi_\alpha}=\bra{\phi_\alpha}$ (i.e. the same as eigenvectors of $H_{\mathrm{w}}$), and
$
\delta_\alpha = -\bbraket{\phi_\alpha|\Gamma|\phi_\alpha}
$
~\cite{Sam73a}.
Let us first work out the static case. Then, $\delta_\alpha=\frac12\nu(|u^{\mathrm{c}}_\alpha|^2+|v^{\mathrm{c}}_{\alpha}|^2)$ with $u^{\mathrm{c}}_\alpha$ and $v^{\mathrm{c}}_\alpha$ evaluated at the contact site. At zero temperature, $\lim_{V\to E_\alpha} \sigma(V)=\sigma_\alpha L(\frac{V-E_\alpha}{\delta_\alpha})$ where $E_\alpha$ is an energy level of the static system, $L(z)=(1+z^2)^{-1}$ is the Lorentzian, and the peak value,
\begin{equation}
\sigma_\alpha = \frac{2e^2}{2\pi}\left|\frac{2 u^{\mathrm{c}}_\alpha v^{\mathrm{c}}_\alpha}{|u^{\mathrm{c}}_\alpha|^2+|v^{\mathrm{c}}_\alpha|^2}\right|^2.
\end{equation}
For the zero-energy Majorana fermion, $u_0=v_0^*$, so $\sigma_0=2e^2/h$ in restored units, as is well known~\cite{LawLeeNg09a,RoyBolSha12a}. In Fig.~\ref{fig:static}, we compare this analytical expression with a full numerical solution. 

\emph{Floquet sum rule.}---In the driven system, the peaks at $V=\epsilon_0$ and $V=\epsilon_\pi$ are not quantized even when Floquet Majorana fermions are present. This is because energies $\epsilon_\alpha+n\Omega$ are all connected via the drive force by emission and absorption processes. Instead, we find a ``Floquet sum rule'' for the sum of differential conductance at these energies~\cite{Note1},
\begin{equation}
\widetilde \sigma(V) = \sum_n \sigma(V+n\Omega).
\end{equation}
At zero temperature, $\lim_{V\to\epsilon_\alpha}\widetilde\sigma(V)=\widetilde\sigma_\alpha L(\frac{V-\epsilon_\alpha}{\delta_\alpha})$ is, again, a Lorentzian with the peak value,
\begin{equation}\label{eq:sigmatilde}
\widetilde\sigma_\alpha = \frac{2e^2}{2\pi}\left|\frac{2\Vert u^{\mathrm{c}}_\alpha\Vert\,\Vert v^{\mathrm{c}}_\alpha\Vert}{\Vert u^{\mathrm{c}}_\alpha \Vert^2 + \Vert v^{\mathrm{c}}_\alpha\Vert^2}\right|^2,
\end{equation}
where $\Vert z \Vert^2 = \sum_k |z^{(k)}|^2$. By particle-hole symmetry $\Vert u_0 \Vert = \Vert v_0 \Vert$ and $\Vert u_\pi \Vert = \Vert v_\pi \Vert$.
Thus, if there is a Floquet Majorana fermion at $\epsilon_0$ and/or $\epsilon_\pi$,
\begin{equation}
\widetilde\sigma_0\equiv\widetilde\sigma(\epsilon_0)= \frac{2e^2}{h}~~\mathrm{and/or}~~\widetilde\sigma_\pi\equiv\widetilde\sigma(\epsilon_\pi) = \frac{2e^2}{h},
\end{equation}
respectively. These relations can be generalized for non-point contact terms as well. This is our central result.

\begin{figure*}[tb]
\includegraphics[width=6.7in]{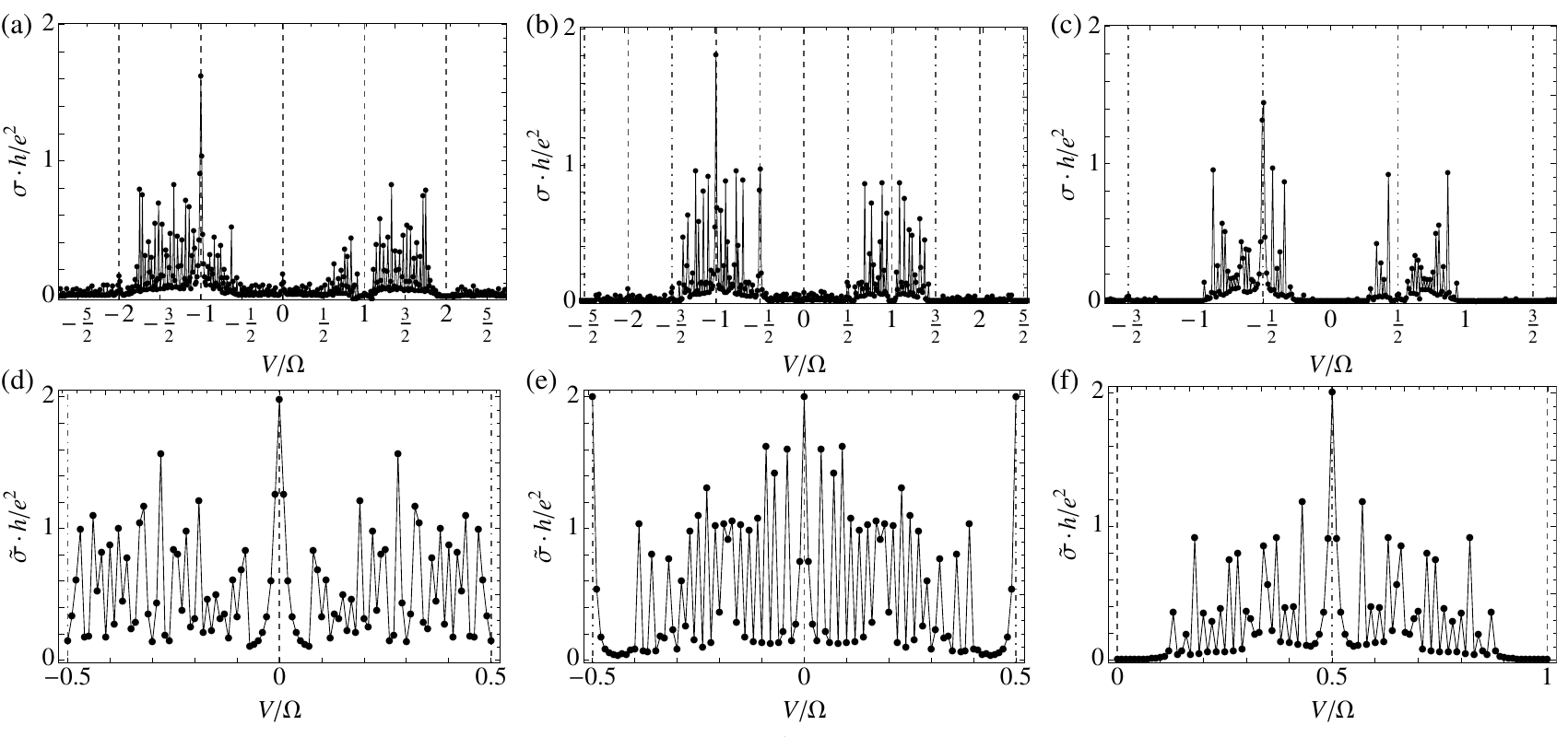}
\caption{Differential conductances $\sigma$ (top row) and $\widetilde\sigma$ (bottom row)  of the driven system as a function of bias $V/\Omega$ in a two-terminal setup. The parameters $\Delta, \mu_1, \mu_2$ and $t_1/t_2$ are as in Fig.~\ref{fig:floqspec}, $\nu/w=2\pi/25$, and the other parameters are: (a,d) $L=40, \Omega/2w=0.37$, (b,e) $L=70, \Omega/2w=0.49$, (c,f) $L=40, \Omega/2w=0.75$. The frequencies are marked by arrows in Fig.~\ref{fig:floqspec}.}\label{fig:floquet}
\end{figure*}

The two Floquet Majorana fermions overlap and split away from $\epsilon_0$ or $\epsilon_\pi$ by an amount $\lambda$ that is exponentially small in their separation. When $\lambda>\nu$, $\widetilde\sigma$ also splits with the peak values $\widetilde\sigma_0$ and $\widetilde\sigma_\pi$ shifting to $V=\epsilon_0\pm\lambda$ and $V=\epsilon_\pi\pm\lambda$, respectively, each with half the widths of the central peak. Therefore, the total weight stays the same. This is a general feature: the total weight $\int_{0}^{\Omega}\widetilde\sigma(V)\dd V=\int_{-\infty}^{\infty}\sigma(V)\dd V \propto\sum_n\int|\mathcal{G}^{(n)}_{\mathrm{eh}}(\omega)|^2\dd\omega$ is constant at all temperatures.

\begin{figure}[t]
\includegraphics[width=2.4in]{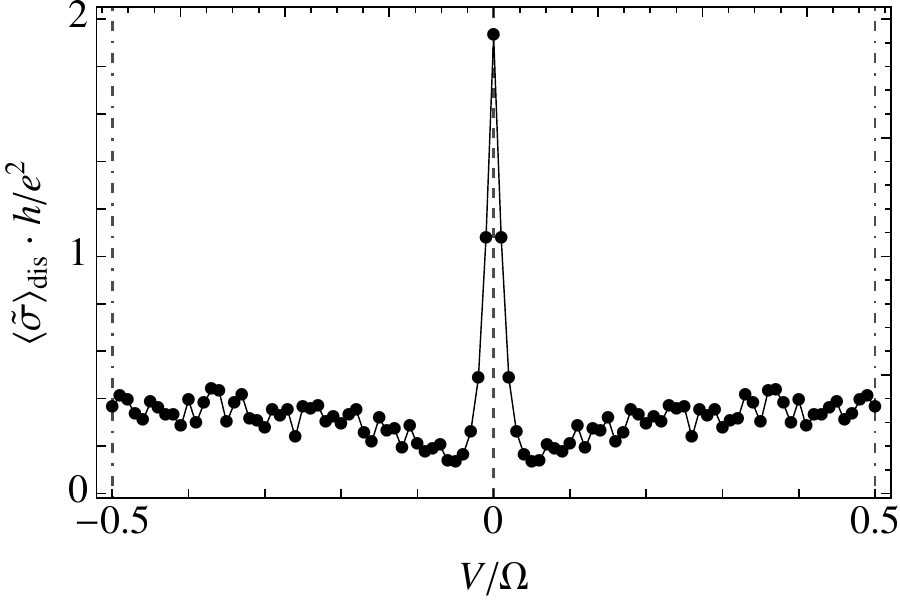}
\caption{Differential conductance in a two-terminal setup averaged over 50 disorder configurations. The parameters are as in Fig.~\ref{fig:floquet} (a,d) and disorder strength $\mu_d=0.28w=0.38\Omega$.}
\label{fig:disorder}
\end{figure}

We have numerically investigated the Floquet sum-rule quantization in the quantum wire. The plots in Fig.~\ref{fig:floquet} show the steady differential conductance calculated for a two-lead setup with symmetric biases $\pm V$. It is clear that, within our numerical precision, $\widetilde\sigma_0$ and/or $\widetilde\sigma_\pi$ are quantized at $2e^2/h$ exactly when $\epsilon_0$ and/or $\epsilon_\pi$ Floquet Majorana fermions appear. Note that the individual peaks of $\sigma$ at $V=n\Omega$ (for $\epsilon_0$ Floquet Majorana fermion) or $V=(2n+1)\Omega/2$ (for $\epsilon_\pi$ Floquet Majorana fermion) are not quantized. Indeed, the main contribution is not even from $n=0$~\cite{Note1}. 
The Floquet spectrum is naturally reflected in $\widetilde\sigma$: The quantized peaks at $\epsilon_0$ (or $\epsilon_\pi$) are separated from the other peaks by a value of $V\sim\epsilon_{g,0}$ (or $\epsilon_{g,\pi}$), i.e. the gap in the quasienergy gap separating the respective Floquet Majorana fermions from the other states. The quasienergy gaps in Fig.~\ref{fig:floquet} are $\sim0.1\Omega$ except for $\epsilon_{g,0}\sim0.05\Omega$ in Fig.~\ref{fig:floquet}(e).

\emph{Effects of disorder.}---The natural question to answer at this point is whether and how $\widetilde\sigma$ could be measured in an actual experiment. It is especially important to be able to tell apart a quantized peak from the other features, which is complicated if $\epsilon_{g,0}$ and $\epsilon_{g,\pi}$ are small. A possible way around is to exploit the topological character of the quantization of $\widetilde\sigma_0$ and $\widetilde\sigma_\pi$. Specifically, they must be protected against disorder while the other features are not. We have studied the effects of disorder numerically by adding a static, uncorrelated, random $\delta\mu_r$ to the wire's chemical potential at site $r$, i.e. $\braket{\delta\mu_r}_{\mathrm{dis}}=0$ and $\braket{\delta\mu_r \delta\mu_{r'}}_{\mathrm{dis}}=\mu_d^2\delta_{rr'}$ where $\mu_d$ is the disorder strength. 
A typical result for the disorder-averaged $\braket{\widetilde\sigma}_{\mathrm{dis}}$ at moderate disorder is shown in Fig.~\ref{fig:disorder}.  The central quantized peak remains nearly unchanged, while the other peaks are suppressed significantly. Note that here $\mu_d>\epsilon_{g,0}$. For stronger disorder the quantized peak is suppressed as well~\cite{Note1}.

\emph{Concluding remarks.}---In sum, we find that the differential conductance summed over periodic drive harmonics, $\widetilde\sigma$, signals Floquet Majorana fermions with a topologically protected quantized value $2e^2/h$ at the Floquet Majorana quasienergy. The quantization is robust and most prominent in the presence of weak disorder. This suggests disorder can be used as a knob to probe Floquet Majorana fermions. At lower frequencies where rotating-wave and similar approximations~\cite{LinRefGal11a,KitOkaBra11a} fail, we have numerically found steady state Floquet Majorana fermions, with similar transport signatures with or without disorder~\cite{Note1}. This is important for possible realization schemes in solid-state systems.
The finite temperature behavior is discussed in the Supplemental Material~\cite{Note1}.

Other transport signatures of Floquet Majorana fermions, such as noise and heat transport, are interesting, open problems. A thorough study of the low-frequency regime is also quite important. Our inclusion of static disorder is appropriate if disorder is intrinsic to the wire itself and not the drive. Other disorder configurations, e.g. in the contacts or the external drive itself, would be interesting to study in future. 
Finally, the effects of disorder at finite temperature as well as interactions are left to future studies.

We acknowledge useful communications with H. Fertig and A. Levchenko. This research is supported by the College of Arts and Sciences at Indiana University, Bloomington.


\vspace{-5mm}


\renewcommand{\thefigure}{S\arabic{figure}}
\setcounter{figure}{0}
\renewcommand{\theequation}{S\arabic{equation}}
\setcounter{equation}{0}

\newpage
\section*{Supplemental Material}
Here we sketch the details for the derivation of the conductance and \textit{Floquet sum rule} described in the main text. We employ Green's function approach for deriving the charge current in the system described by time dependent Hamiltonian $H_{\mathrm{w}}(t)$ and contact Hamiltonian $H_{\mathrm{c}}$, as discussed in the main text
\begin{align}
H_{\mathrm{w}}(t) = \frac{\ii}2 \gamma^{\intercal}A(t)\gamma, \quad H_{\mathrm{c}}=\sum_{\lambda}a^{\lambda\dagger}K^{\lambda}\gamma+\mathrm{h.c.}
\end{align}
The net charge current flowing across the contact $\lambda$ into the wire is ($\hbar=1$)
\begin{align}
 J^{\lambda} (t) &= \ii e\left[H_{\mathrm{w}}(t)+H_{\mathrm{c}},N^{\lambda}(t)\right] \nonumber \\
 &= \ii e\left( \gamma^{\intercal}(t)K^{\lambda\dagger}a^{\lambda}(t) - \mathrm{h.c.} \right),\label{spl:curr-def}
\end{align}
where $N^\lambda$ is the number operator for electrons in lead $\lambda$. One can use the solution of Heisenberg' equation for the electron (in the lead)  $a^{\lambda} (t)$ 
\begin{align}\label{spl:at}
 a^{\lambda}(t) = \eta^{\lambda}(t) + \int_{t_0 \rightarrow -\infty}^t g^{\lambda} (t-t') K^{\lambda} \gamma (t') \dd t',
\end{align}
where, $t_0$ is the \textit{switching time}, $g^{\lambda} (t-t')$ is the Green's function matrix of electrons in lead $\lambda$. In the wide band limit the density of states $\rho^{\lambda}$ of lead $\lambda$ is constant for the relevant energy scales and in the simplest situation $g^{\lambda}(\omega) = -i\pi \rho^{\lambda} $. The \textit{noise} term  $\eta^{\lambda} (t) = \ii g^{\lambda} (t-t_0)a^{\lambda} (t_0)$ obeys the fluctuation-dissipation relation after averaging over the lead states
\begin{align}
 &\langle \eta^{\lambda \dagger}_r(\omega)\eta^{\lambda'}_{r'}(\omega') \rangle = (2\pi)^2 \delta_{\lambda \lambda'}\rho^{\lambda}_{rr'}f_{\lambda}(\omega)\delta(\omega - \omega'), \\
 &\langle \eta^{\lambda}_r(\omega)\eta^{\lambda'\dagger}_{r'}(\omega') \rangle = (2\pi)^2 \delta_{\lambda \lambda'}\rho^{\lambda}_{rr'}\bar f_{\lambda}(\omega)\delta(\omega - \omega'),
\end{align}
where $f_{\lambda}(\omega)= 1-\bar f_\lambda(\omega) = \left[1+e^{(\omega-eV_{\lambda})/\tau_{\lambda}}\right]^{-1}$ is the Fermi-Dirac distribution of the  lead $\lambda$ with bias $V_{\lambda}$ and temperature $\tau_{\lambda}$. (The Boltzmann constant $k_B=1$.) For the Majorana operator $\gamma(t)$, the integration of the Heisenberg equation can be complicated by the time dependence of $A(t)$. In a periodically driven system it is obtained in terms of the Floquet Green's function, as we discuss later.

\subsection{Static System}
If the time dependence of $A(t)$ is trivial, one can integrate the Heisenberg's equation for the Majorana operator with using Eq.~(\ref{spl:at}). In the Fourier space,
\begin{align}\label{spl:gamma-s}
 \gamma(\omega) = G(\omega) h(\omega),
\end{align}
where $G(\omega)$ is the Green's function defined from
\begin{align}
G^{-1}(\omega) = \omega  - \ii A - \ii \Gamma(\omega),
\end{align}
with the self energy \begin{align}
\ii\Gamma(\omega) = \left[ K^{\lambda \dagger} g^{\lambda}(\omega) K^{\lambda} - {K^{\lambda}}^\intercal g^{\lambda *}(-\omega) K^{\lambda *} \right]
\end{align}
and 
\begin{align}
h(\omega)  = \sum_{\lambda} \left[ K^{\lambda \dagger} \eta^{\lambda}(\omega) - {K^{\lambda}}^\intercal \eta^{\lambda *}(-\omega)\right]. 
\end{align}

Using Eq.~(\ref{spl:gamma-s}) and Eq.~(\ref{spl:at}) in the current equation Eq.~(\ref{spl:curr-def}) and averaging over the lead states one obtains~\cite{roy2012}
\begin{widetext}
\begin{align}
 J^{\lambda}= \frac{e}{2\pi } \sum_{\lambda'} \int \dd\omega\left\{ 
 \mathrm{Tr}\left[ \xi^{\lambda}G(\omega)\xi^{\lambda'}G^{\dagger}(\omega) \right]\left[f_{\lambda'}(\omega) -f_{\lambda}(\omega) \right] + \mathrm{Tr}\left[ \xi^{\lambda}G(\omega){\xi^{\lambda'}}^\intercal G^{\dagger}(\omega) \right] \left[1-f_{\lambda'}(-\omega) -f_{\lambda}(\omega) \right]\right\},
\end{align}
where $\xi^{\lambda} = 2\pi K^{\lambda\dagger}\rho^{\lambda}K^{\lambda}$ and $J^{\lambda} = \braket{J^\lambda(t)}$ (the average is over the lead states). The conductance is obtained by taking the derivative with respect to the bias $V_{\lambda}$,
\begin{align}
 \sigma^{\lambda} = -\frac{e^2}{2\pi}\int \dd\omega\mathrm{Tr}\left[ \xi^{\lambda}G(\omega){\xi^{\lambda}}^\intercal G^{\dagger}(\omega) \right]\left[f'_{\lambda}(-\omega) +f'_{\lambda}(\omega) \right]  -\frac{e^2}{2\pi} \sum_{\lambda'\neq \lambda}\int \dd\omega\mathrm{Tr}\left[ \xi^{\lambda}G(\omega)(\xi^{\lambda'}+{\xi^{\lambda'}}^\intercal)G^{\dagger}(\omega) \right]f'_{\lambda}(\omega).
\end{align}
\end{widetext}
The second term vanishes if the system has a single lead,
\begin{align}\label{spl:cond}
 \sigma^{\lambda} = -\frac{e^2}{2\pi}\int \dd\omega~&\mathrm{Tr}\left[ \xi^{\lambda}G(\omega){\xi^{\lambda}}^\intercal G^{\dagger}(\omega) \right]\left[f'_{\lambda}(-\omega) +f'_{\lambda}(\omega) \right].
\end{align}

\begin{figure}
\centering
\includegraphics[width=2.8in]{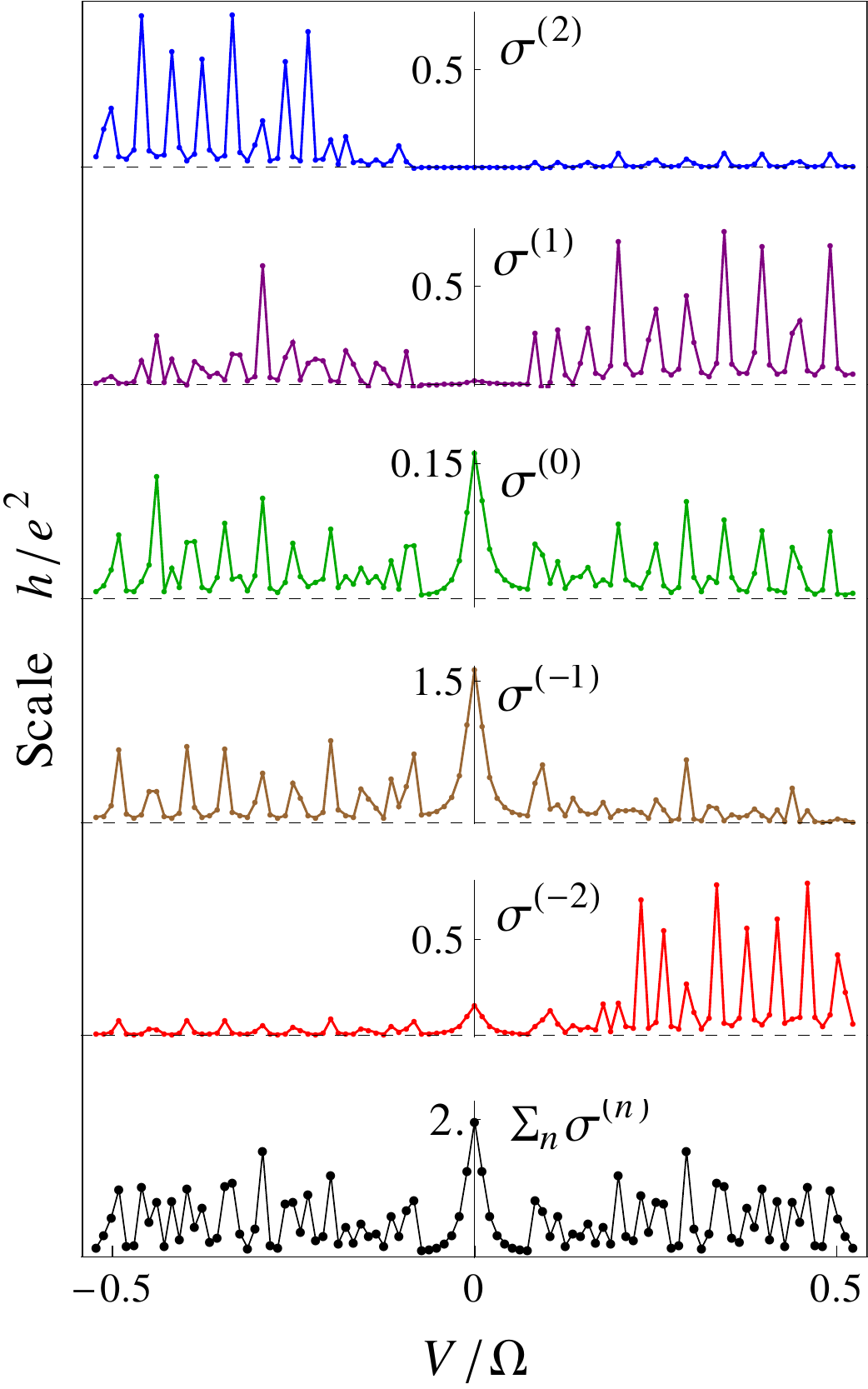}
\caption{Contributions to Floquet sum rule from different $n$ in $\widetilde{\sigma}(V)=\sum_n\sigma(n\Omega+V) $. Here $\sigma^{(n)}=\sigma(n\Omega+V)$. The parameters are as in Fig. 3 (c,f) of the main text.}
\label{spl-fig:sum}
\end{figure}

\begin{figure*}[t]
\centering
\includegraphics[width=1.01\textwidth]{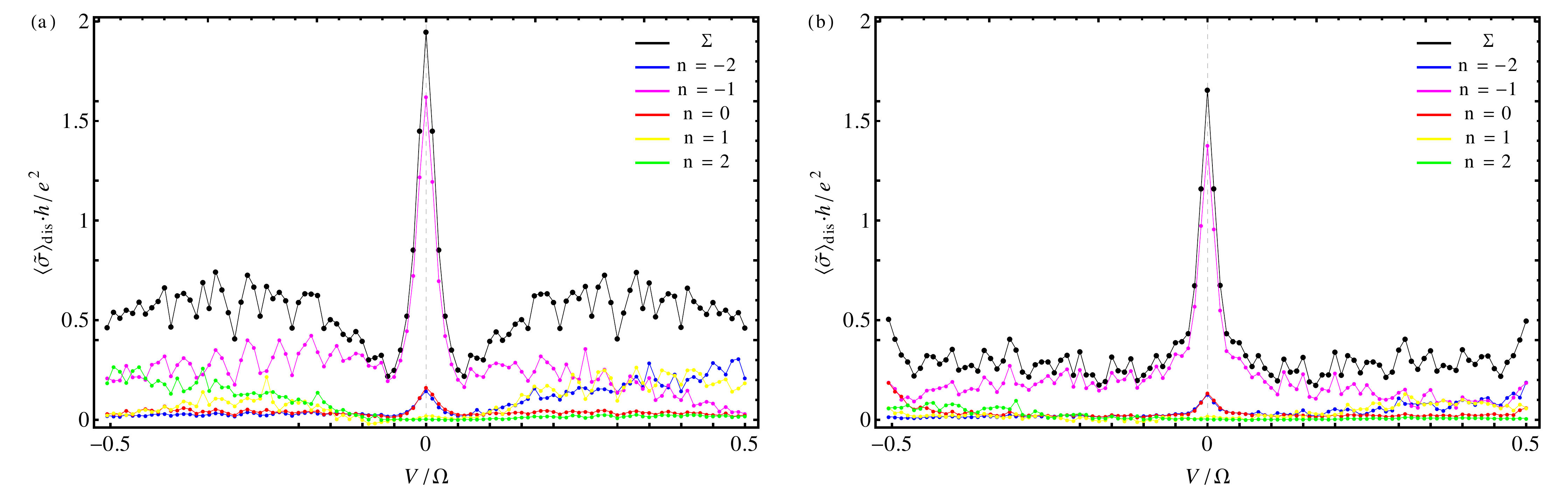}
\caption{Differential conductance in a two-terminal setup averaged over 50 disorder configurations, where  disorder strengths (a) $\mu_d=0.2 w=0.26\Omega$ and (b) $\mu_d=0.44 w=0.6 \Omega$ which are smaller and larger than the disorder strength used in the main text. Other parameters are as in Fig.~\ref{spl-fig:sum}, except here $\nu/w= 2\pi/17$. We show the contribution to the sum, $\widetilde{\sigma}(V) \equiv \Sigma$, from different $\sigma^{(n)}(V) = \sigma(V+n\Omega)$.}\label{fig:other-dis}
\end{figure*}

\subsection{Periodically driven system}
In a driven system, the formulation of current by non-equilibrium Floquet Green's function has been discussed before~\cite{arrachea2002,arrachea2005,arrachea2006} and we follow the same strategy. Integrating the Heisenberg equation for the Majorana operator in a driven system gives~\cite{kohler2005}
\begin{align}
 \left[\ii \frac{\partial}{\partial t}  - \ii A(t) \right]\gamma(t) - \ii \int_0^{\infty}\dd s~\Gamma(s)\gamma(t-s) = h(t),
\end{align}
with self energy  $\ii\Gamma (s) =2i\sum_{\lambda}\mathrm{Im}\left[ K^{\lambda \dagger} g^{\lambda}(s) K^{\lambda} \right]$ and $h(t)  = 2\ii\sum_{\lambda}\mathrm{Im} \left[ K^{\lambda \dagger} \eta^{\lambda}(t) \right]$. 
The Green's function of this inhomogeneous equation $G(t,t')$ satisfies
\begin{align}\label{spl:GFF}
 \left[\ii \frac{\partial}{\partial t}  - \ii A(t) \right]G(t,t') - &\ii \int_0^{\infty}\dd s~\Gamma(s)G(t-s,t')\dd s \nonumber \\
 &= \delta(t-t').
\end{align}
For a periodic drive with period $T = 2\pi/\Omega$, the \textit{Floquet Green's function} is also periodic over the same period $G(t+T,t'+T) = G (t,t')$. One can introduce the Fourier transform,
\begin{align}
G(t,t') &= \sum_k\int \frac{\dd\omega}{2\pi}~ G^{(k)}(\omega) e^{-\ii\omega(t-t')}e^{-\ii k\Omega t},
\end{align}
The Majorana operator is solved in terms of the Green's function
\begin{align}
 \gamma(t) = \sum_k \int \frac{\dd\omega}{2\pi} e^{-\ii\omega t}e^{-\ii k\Omega t}G^{(k)}(\omega)h(\omega).
\end{align}
Using these expressions in the current formula  Eq.~(\ref{spl:curr-def}), the the steady state current $J^\lambda = \bbraket{J^\lambda} \equiv \int_0^T \braket{J^\lambda(t)} \dd t/T$ in the wide band limit is found to be
\begin{widetext}
 \begin{align}
 J^{\lambda} =& -\frac{e}{2\pi }\sum_{\lambda'}\int \dd\omega ~\sum_l \Bigg\{\mathrm{Tr}\left[G^{(l)\dagger}(\omega)\left(\xi^{\lambda'}+ {\xi^{\lambda'}}^\intercal\right)G^{(l)}(\omega)\xi^{\lambda} \right]f_{\lambda}(\omega) \nonumber \\
 & -\mathrm{Tr}\left[\xi^{\lambda}G^{(l)}(\omega)\xi^{\lambda'}G^{(l)\dagger}(\omega) \right]f_{\lambda'}(\omega) - \mathrm{Tr}\left[\xi^{\lambda}G^{(l)}(\omega){\xi^{\lambda'}}^\intercal G^{(l)\dagger}(\omega) \right]\left[1-f_{\lambda'}(-\omega)\right]\Bigg\},
\end{align}
which gives the conductance
\begin{align}\label{spl:condFull}
 \sigma^{\lambda}= -\frac{e^2}{2\pi}\int \dd\omega~ \sum_l &\mathrm{Tr}\left[ \xi^{\lambda}G^{(l)}(\omega){\xi^{\lambda}}^\intercal G^{{(l)}\dagger}(\omega) \right]\left[f'_{\lambda}(-\omega) +f'_{\lambda}(\omega) \right] \nonumber \\
& -\frac{e^2}{2\pi} \sum_{\lambda'\neq \lambda}\int \dd\omega~\sum_l\mathrm{Tr}\left[ \xi^{\lambda}G^{(l)}(\omega)(\xi^{\lambda'}+{\xi^{\lambda'}}^\intercal)G^{{(l)}\dagger}(\omega) \right]f'_{\lambda}(\omega) .
\end{align}
\end{widetext}
Finally, similar to the static problem, for a single lead, 
\begin{align}\label{spl:condF}
 \sigma^{\lambda} = -\frac{e^2}{2\pi}\sum_l \int \dd\omega\, &\mathrm{Tr}\left[ \xi^{\lambda}G^{(l)}(\omega){\xi^{\lambda}}^\intercal G^{{(l)}\dagger}(\omega) \right]\nonumber \\
& \times \left[f'_{\lambda}(-\omega) +f'_{\lambda}(\omega) \right].
\end{align}

\subsection{Relation to Nambu-Gorkov Green's function}
The Majorana representation is related to the Nambu spinor $(c_r^{\phantom{\dagger}},c_r^{\dagger})=\psi_r^{\intercal}$  by the unitary transformation $\gamma_{r}=u\psi_r$ with $u=\frac1{\sqrt 2}\left(\begin{array}{cc} 1 & 1 \\ -\ii & \ii\end{array}\right)$ for each site $r$. And the Bogoliubov-de Gennes (BdG) Hamiltonian is related as $H_{\mathrm{BdG}}=\ii U^{\dagger} A U$, where $U=u\oplus u\oplus \dots \oplus u$ ($L$ times). The Nambu-Gorkov Green's function in a static system is defined as
\begin{align}
 \mathcal{G} (\omega) = \left(\omega -H_{\mathrm{BdG}} +i\epsilon \right)^{-1} ,
\end{align}
with vanishing positive $\epsilon$. 

For a single lead with point contact, the matrix $\xi = \nu (1-\sigma_y)/2$. By unitary transforming the Eq.~(\ref{spl:cond}) with $U$ we get
\begin{align}
 \sigma = -\frac{e^2\nu^2}{2\pi} \int \dd\omega~\left|\mathcal{G}_{\mathrm{eh}}(\omega) \right|^2\left[ f'(-\omega) +f'(\omega) \right],
\end{align}
where $\mathcal{G}_{\mathrm{eh}}$ is the electron-hole component of the Green's function evaluated at the contact site
\begin{align}
\mathcal{G}_{\mathrm{c}}(\omega)= \left(\begin{array}{cc}\mathcal{G}_{\mathrm{ee}}(\omega) & \mathcal{G}_{\mathrm{eh}}(\omega)\\                                                                                    
\mathcal{G}_{\mathrm{he}}(\omega) &\mathcal{G}_{\mathrm{hh}}(\omega)                                                                                           \end{array}\right)
,
\end{align}
For the periodically driven system the relevant Green's function is the Floquet Green's function in Nambu spinor basis, which is defined as~\cite{kohler2005}
\begin{align}
 \mathcal{G}^{(l)}(\omega) = \sum_{\alpha k} \frac{\ket{\varphi_\alpha^{(l+k)}}\bra{\bar{\varphi}_\alpha^{(k)}}}{\omega-\epsilon_\alpha-l\Omega+\ii \delta_\alpha},
\end{align}
where $\ket{\varphi_\alpha^{(i)}}$ and $\bra{\bar{\varphi}_\alpha^{(k)}}$ are the discrete Fourier component of the time periodic function 'Floquet states' $\ket{\varphi_\alpha(t)}$ and $\bra{\bar{\varphi}_\alpha(t)}$, which are right and left eigenstates of the effective Floquet Hamiltonian
$H_{\mathrm{eff}}=H_{\mathrm{w}}+\ii \Gamma -\ii \frac{\partial}{\partial t}$ with eigenvalues (\textit{Floquet energies}) $\epsilon_\alpha - \ii \delta_{\alpha}$.
\begin{align}
 H_{\mathrm{eff}}\ket{\varphi_\alpha(t)} &= (\epsilon_{\alpha}-\ii \delta_{\alpha})\ket{\varphi_\alpha(t)}, \\
 \bra{\bar{\varphi}_\alpha(t)}H_{\mathrm{eff}} &=(\epsilon_{\alpha}-\ii \delta_{\alpha}) \bra{\bar{\varphi}_\alpha(t)}. \nonumber
\end{align}
$\ket{\varphi_\alpha(t)}$ lives in an extended (with the time axis) Hilbert space~\cite{Sam73a-Sup}, where the expectation of any operator $\mathcal{O}$ in the state is defined with the time average
\begin{align}
 \bbraket{\mathcal{O}}=\frac1T\int_0^T \mathrm{d}t~\bra{\bar{\varphi}_\alpha(t)}\mathcal{O}\ket{\varphi_\alpha(t)} =\sum_k\bra{\bar{\varphi}_\alpha^{(k)}}\mathcal{O}\ket{\varphi_\alpha^{(k)}}.
\end{align}

In terms of the Green's functions $\mathcal{G}^{(l)}$, one can express the conductance for a periodically driven system with single lead and contact matrix $\xi = \nu (1-\sigma_y)/2$ similar to the static case,
\begin{align}
 \sigma = -\frac{e^2\nu^2}{2\pi}\sum_n\int\left[|\mathcal{G}_{\mathrm{he}}^{(n)}(\omega)|^2+|\mathcal{G}_{\mathrm{eh}}^{(n)}(-\omega)|^2\right]f'(\omega)\dd\omega,
\end{align}

Now, the eigenstate of the time dependent Hamiltonian $H_{\mathrm{w}}$, $\ket{\psi_{\alpha}(t)}$ are related to the Floquet state $\ket{\varphi_{\alpha}(t)}$ by the relation $\ket{\psi_{\alpha}(t)} = e^{-\ii(\epsilon_{\alpha}-\ii\delta_{\alpha})t}\ket{\varphi_{\alpha}(t)}$.
For 0 and $\pi/T$ Floquet energy, $\ket{\psi_0(t)}=e^{-\delta_{0}t}\ket{\varphi_0(t)}$ and $\ket{\psi_{\pi}(t)}=e^{-\ii \pi t/T}e^{-\delta_{\pi}t}\ket{\varphi_{\pi}(t)}$. Considering the contact term perturbatively, in the leading order one can neglect the modification to the wavefunction to have 
$\ket{\psi_0(t)}\approx \ket{\varphi_0(t)}$ and $\ket{\psi_{\pi}(t)}\approx e^{-\ii \pi t/T}\ket{\varphi_{\pi}(t)}$. Using the particle-hole symmetry at Floquet energies $0,~\pi/T$, one can express the eigenstates in the Bogoliubov quasiparticle representation $\ket{\psi_0(t)}=(u_0(t),u_0^*(t))^\intercal$ and $\ket{\psi_{\pi}(t)}=(u_{\pi}(t),u_{\pi}^*(t))^\intercal$. This allows us to write $\ket{\varphi_i(t)}$ in the Fourier space
\begin{align}
 \ket{\varphi_{0}^{(l)}} = \left(\begin{array}{l}
                           u_0^{(l)} \\
                           u_0^{(-l)*}
                          \end{array}\right),~~~~ \ket{\varphi_{\pi}^{(l)}} = \left(\begin{array}{l}
                           u_0^{(l)} \\
                           u_0^{(-l+1)*}
                          \end{array}\right).
\end{align}
The leading perturbation to the self-energy part can be computed in the extended Hilbert space~\cite{Sam73a-Sup} with 
\begin{equation}
\Gamma = -(\xi+\xi^{\intercal})/2= -\frac{\nu}2 \left( \begin{array}{cc} 1 & 0 \\ 0 & 1 \end{array}\right)_{\mathrm{c}}\oplus\underbrace{0\oplus 0  ...\oplus 0}_{L-1~\mathrm{times}},
\end{equation} 
\begin{align}
 \delta_{\alpha} = -\bbra{\varphi_{\alpha}(t)}\Gamma\kket{\varphi_{\alpha}(t) }= \frac{\nu}2 (||u_{\alpha}^{\mathrm{c}}||^2+||v_{\alpha}^{\mathrm{c}}||^2),
\end{align}
where $u_{\alpha}^{\mathrm{c}(k)}$ and $v_{\alpha}^{\mathrm{c}(k)}$ are evaluated at the contact site and $||z||^2=\sum_k|z^{(k)}|^2$. We also note that 
\begin{align}
\lim_{V\rightarrow \epsilon_{\alpha}} & \sum_k|\mathcal{G}_{\mathrm{eh}}^{(l)}(V + k\Omega)|^2 \approx \sum_{lk}\left|\frac{u_{\alpha}^{\mathrm{c}(k+l)}v_{\alpha}^{\mathrm{c}(k)}}{V-\epsilon_\alpha+\ii \delta_{\alpha}}\right|^2 \nonumber \\
&= \frac{4}{|\nu|^2}\left|\frac{||u_{\alpha}^{\mathrm{c}}||~ ||v_{\alpha}^{\mathrm{c}}||}{||u_{\alpha}^{\mathrm{c}}||^2 + ||v_{\alpha}^{\mathrm{c}}||^2}\right|^2 L\left(\frac{V-\epsilon_\alpha}{\delta_\alpha}\right),
\end{align}
where $L(z)=(1+z^2)^{-1}$ is the Lorentzian. The peak value at zero temperature is
\begin{align}
 \widetilde{\sigma}_{\alpha}=\lim_{V\rightarrow\epsilon_{\alpha}}\sum_k\sigma(V+k\Omega)=\frac{2e^2}{2\pi}\left|\frac{2||u_{\alpha}^{\mathrm{c}}||~ ||v_{\alpha}^{\mathrm{c}}||}{||u_{\alpha}^{\mathrm{c}}||^2 + ||v_{\alpha}^{\mathrm{c}}||^2}\right|^2.
\end{align}
This is our Eq.~(9) in the main text.

\begin{figure}[t]
\includegraphics[width=2.8in]{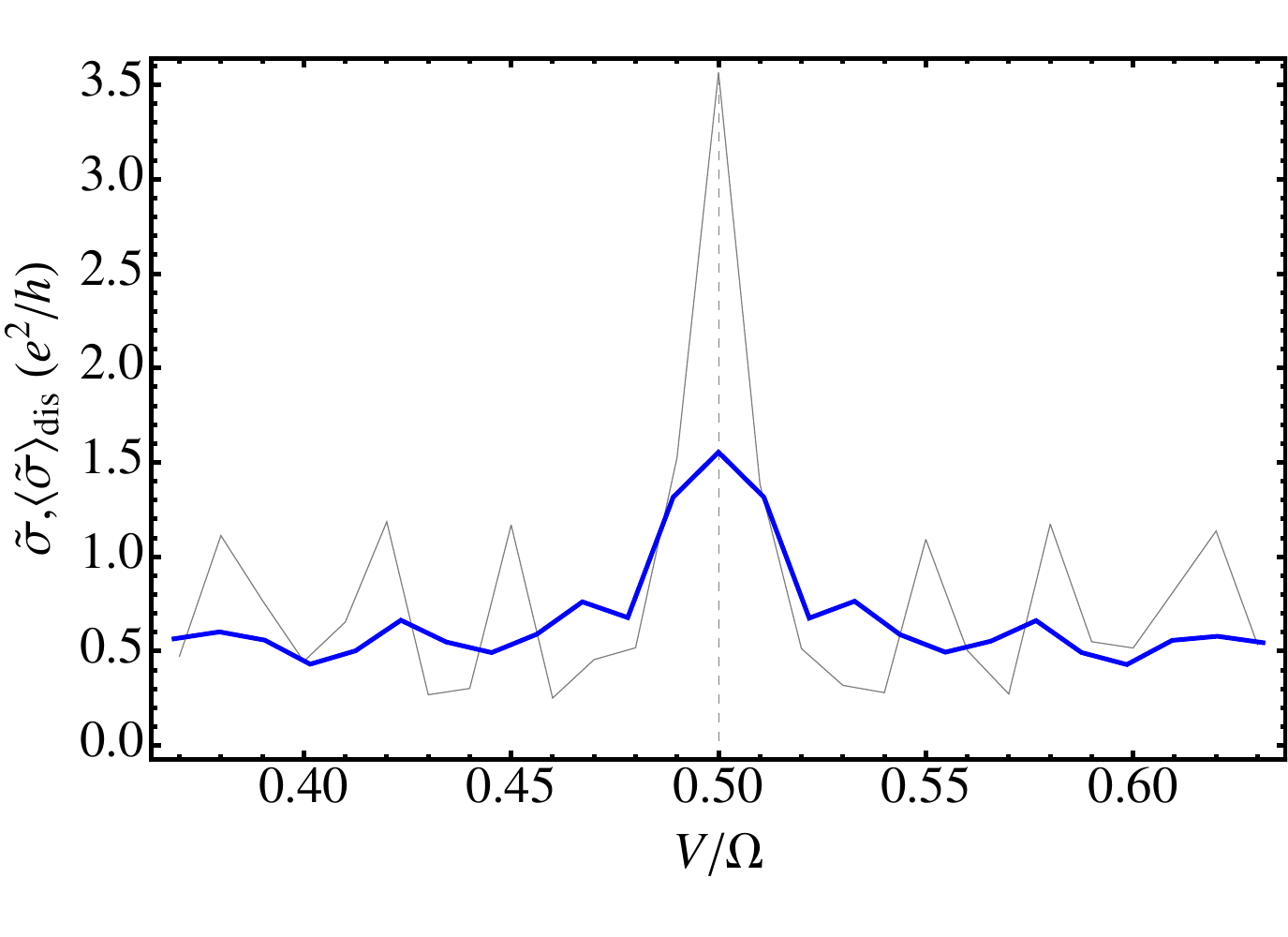}\\
\vspace{-1mm}
\includegraphics[width=2.8in]{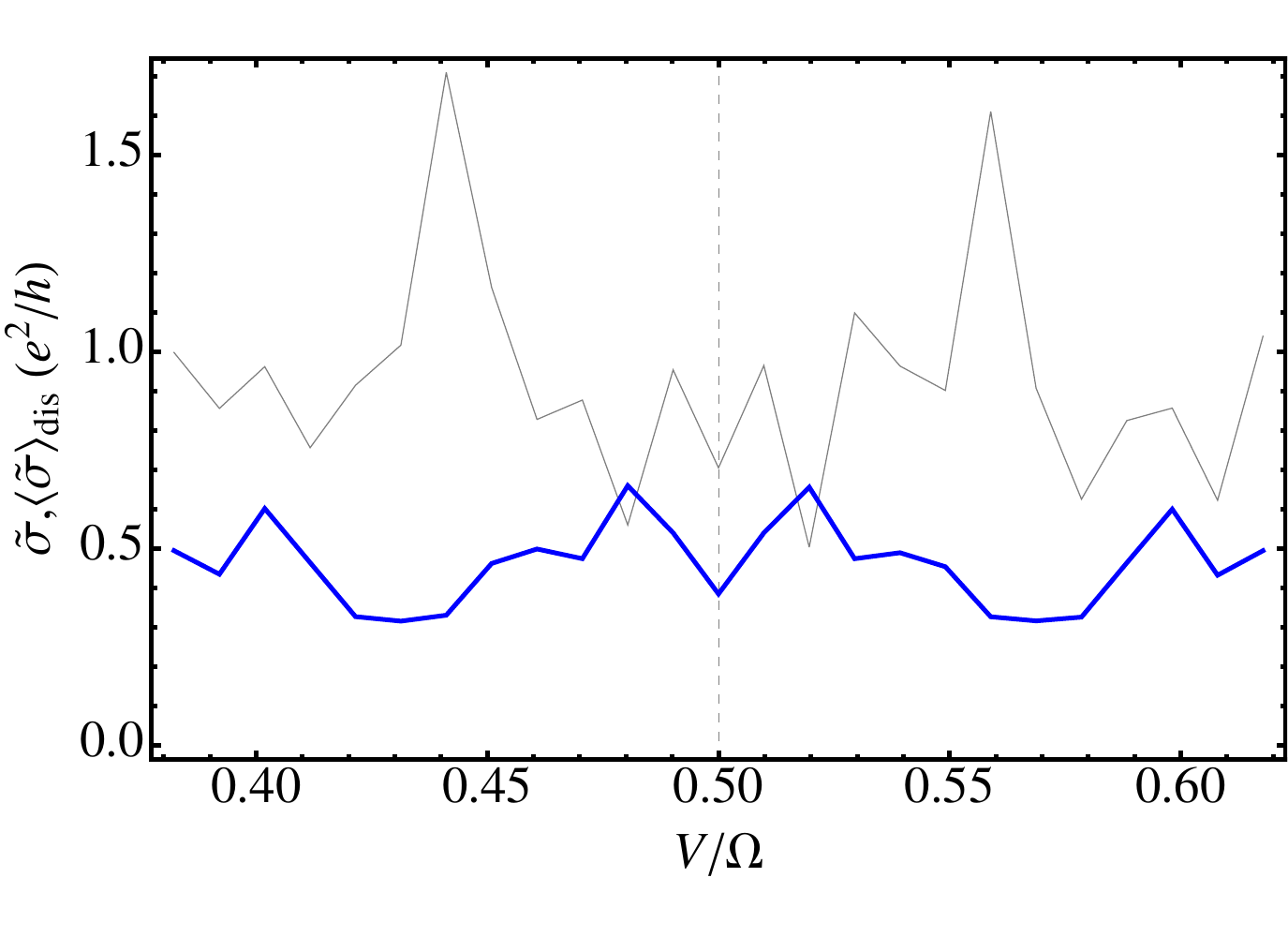}
\caption{Differential conductance near $V=\Omega/2=\epsilon_\pi$ at $\Omega/2w=0.095$ (top) and $\Omega/2w=0.106$ (bottom) for a clean (thin gray) and a disordered (thick blue) system with $\mu_d=0.28w$ averaged over 25 (top) and 20 (bottom) disorder configurations. Other parameters are as in Fig. (1) in the main text, except $L=200$, and $\nu=\pi/25$. There are an odd (5) number of localized modes with quasienergies near $\epsilon_\pi$ in the top panel and an even (4) number in the bottom panel.}\label{fig:lowF}
\vspace{2mm}
\includegraphics[width=2.5in]{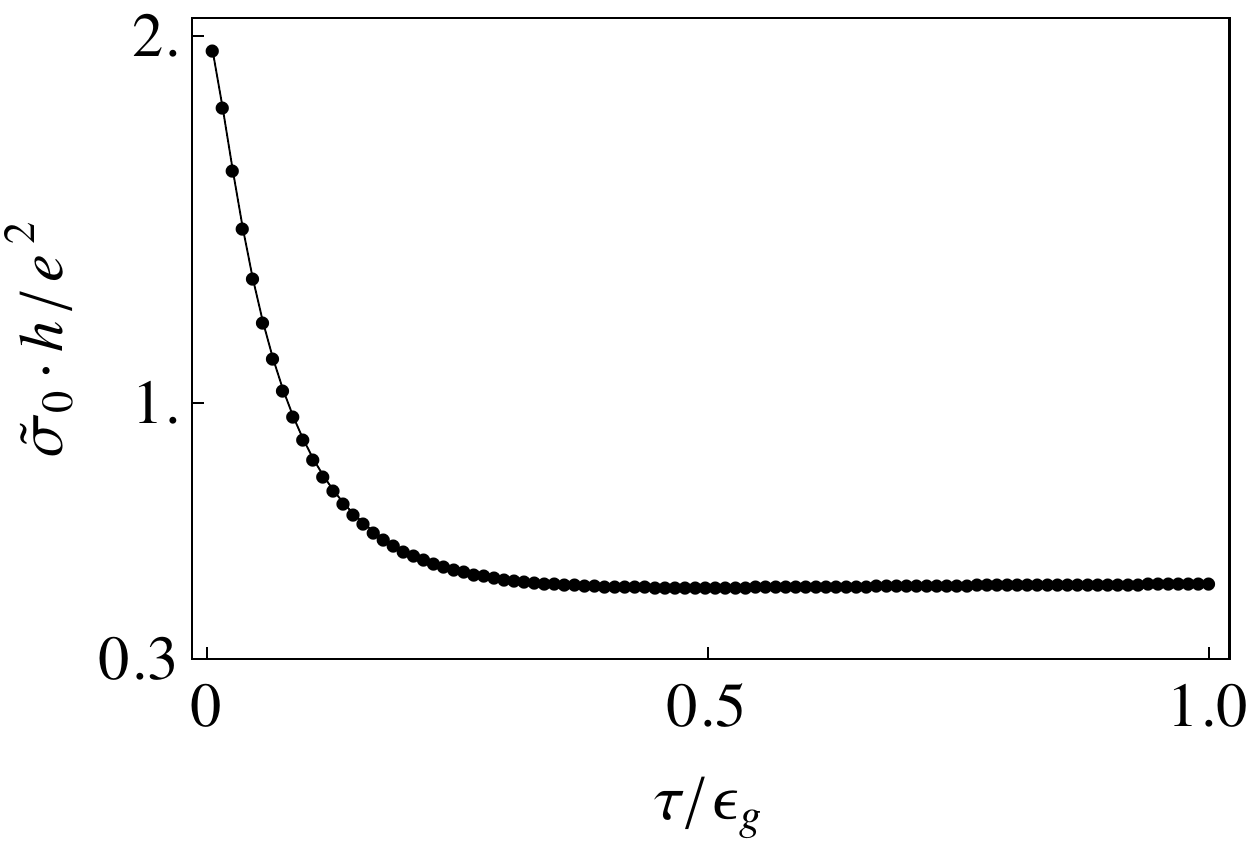}
\caption{Temperature ($\tau$) dependence of the peak height of conductance $\widetilde{\sigma}(0) \equiv\sigma_0 $ for Floquet Majorana. The parameters used are same as that of Fig.~\ref{spl-fig:sum}. The Floquet gap is $\epsilon_g=0.3\Omega$.}\label{fig:finT}
\end{figure}

\vspace{-4mm}
\subsection{Numerical Results}
In Fig.~\ref{spl-fig:sum} we plot $\widetilde{\sigma}(V)=\sum_n\sigma(V+n\Omega)$ for system with two symmetrically biased leads ($V_1=-V_2=-V$) hosting a $\epsilon_0$ Floquet Majorana fermion, which shows that the peaks of individual components $\sigma(V+n\Omega)$ are not quantized, but the sum $\widetilde{\sigma}$ shows a quantized peak. 

In Fig.~\ref{fig:other-dis}, we compare the differential conductance $\widetilde{\sigma}$ for smaller and larger than optimal disorder strength. For small disorder, the non-topological peaks do not get suppressed enough compared to the topological peak, whereas for large disorder the topological peak can get suppressed. 

\vspace{-5mm}
\subsection{Low-frequency limit}
At frequencies much smaller than the bandwidth, the Floquet spectrum may have many quasi-energies close to $\epsilon_0$ and $\epsilon_\pi$ in addition to the Floquet Majorana fermions. In a clean system these non-topological steady states may contribute to steady-state Andreev processes. However, these near $\epsilon_0$ or $\epsilon_\pi$ states are paired at the same edge and will be mixed by disorder. Therefore, in a disordered system, they are predominantly particle-like or hole-like and do not contribute significantly to Andreev reflection amplitude in Eq. (9) $\sim \Vert u_L \Vert \Vert v_L \Vert$. By contrast, an unpaired Floquet Majorana fermion at one end of the wire has its partner localized at the other end. Disorder has little effect in mixing these spatially separated states and the Andreev reflection amplitude remains finite and large.

We verify this picture numerically at drive frequencies $\sim$ 5\% of the bandwidth for cases with an odd or even number of localized modes near $\epsilon_\pi$, with and without disorder. As seen in Fig.~\ref{fig:lowF}, the existence (absence) of a Floquet Majorana fermion in the odd- (even-) number case is signaled by a near-quantized peak (dip) in the presence of disorder.

\vspace{-2mm}
\subsection{Temperature dependence}
Finally, in Fig.~\ref{fig:finT} we plot the temperature dependence of $\widetilde{\sigma}(0)$ with temperature ($\tau$). For $\nu\ll\tau\lesssim\epsilon_g$, the Majorana peak height falls $\sim 1/\tau$. For $\tau\gtrsim\epsilon_g$, the peaks start to overlap and $\widetilde\sigma_{0,\pi}$ saturates.

\end{document}